\documentclass[journal]{IEEEtran}

\ifCLASSINFOpdf
\else
\usepackage[dvips]{graphicx}
\fi
\usepackage{url}
\usepackage{xcolor}
\usepackage{scalerel}
\usepackage{gensymb}
\usepackage{tikz}
\usepackage{enumitem}
\usetikzlibrary{svg.path}

\definecolor{orcidlogocol}{HTML}{A6CE39}
\tikzset{
	orcidlogo/.pic={
		\fill[orcidlogocol] svg{M256,128c0,70.7-57.3,128-128,128C57.3,256,0,198.7,0,128C0,57.3,57.3,0,128,0C198.7,0,256,57.3,256,128z};
		\fill[white] svg{M86.3,186.2H70.9V79.1h15.4v48.4V186.2z}
		svg{M108.9,79.1h41.6c39.6,0,57,28.3,57,53.6c0,27.5-21.5,53.6-56.8,53.6h-41.8V79.1z M124.3,172.4h24.5c34.9,0,42.9-26.5,42.9-39.7c0-21.5-13.7-39.7-43.7-39.7h-23.7V172.4z}
		svg{M88.7,56.8c0,5.5-4.5,10.1-10.1,10.1c-5.6,0-10.1-4.6-10.1-10.1c0-5.6,4.5-10.1,10.1-10.1C84.2,46.7,88.7,51.3,88.7,56.8z};
	}
}

\newcommand\orcidicon[1]{\href{https://orcid.org/#1}{\mbox{\scalerel*{
				\begin{tikzpicture}[yscale=-1,transform shape]
					\pic{orcidlogo};
				\end{tikzpicture}
			}{|}}}}

\usepackage[hidelinks]{hyperref}
\usepackage{amsmath,amssymb,amsfonts}
\usepackage{cite}
\usepackage{graphicx}
\usepackage{tabularx, ragged2e}
\usepackage{dblfloatfix}
\usepackage{arydshln}

\definecolor{net_yellow}{RGB}{254,229,59}
\definecolor{net_blue}{RGB}{29,51,120}
\definecolor{sigmoid_blue}{RGB}{92,205,220}

\begin{document}
	\bstctlcite{BSTcontrol}
	\title{Improving Super-Resolution Performance using Meta-Attention Layers}

	\author{Matthew Aquilina\textsuperscript{\orcidicon{0000-0002-4039-1398}}, Christian Galea\textsuperscript{\orcidicon{0000-0002-0353-7028}}, \textit{Member IEEE}, John Abela\textsuperscript{\orcidicon{0000-0002-8162-5816}}, \textit{Member IEEE}, Kenneth P. Camilleri\textsuperscript{\orcidicon{0000-0003-0436-6408}}, \textit{Senior Member IEEE}, Reuben A. Farrugia\textsuperscript{\orcidicon{0000-0001-8106-9891}}, \textit{Senior Member, IEEE}
		\thanks{The research work disclosed in this publication is funded by MCST Grant number R\&I-2017-002-T.  \textit{(Corresponding author: Matthew Aquilina.)} }
		\thanks{Matthew Aquilina (\href{mailto:matthew.aquilina@um.edu.mt}{matthew.aquilina@um.edu.mt}) is with the Dept. of Communications \& Computer Engineering, Faculty of ICT, University of Malta, Msida, Malta (UM) and the Deanery of Molecular, Genetic \& Population Health Sciences, University of Edinburgh, Edinburgh, Scotland, UK.
			
			Christian Galea (\href{mailto:christian.p.galea@um.edu.mt}{christian.p.galea@um.edu.mt}) and Reuben A. Farrugia (\href{mailto:reuben.farrugia@um.edu.mt}{reuben.farrugia@um.edu.mt}) are with the Dept. of Communications \& Computer Engineering, Faculty of ICT, UM.
			
			John Abela (\href{mailto:john.abela@um.edu.mt}{john.abela@um.edu.mt}) is with the Dept. of Computer Information Systems, Faculty of ICT, UM.
			
			Kenneth P. Camilleri (\href{mailto:kenneth.camilleri@um.edu.mt}{kenneth.camilleri@um.edu.mt}) is with the Dept. of Systems \& Control Engineering, Faculty of Engineering, UM.}
		\vspace{-1cm}
	}

	\markboth{IEEE SIGNAL PROCESSING LETTERS}{}
	\IEEEoverridecommandlockouts
	\IEEEpubid{\makebox[\columnwidth]{1070-9908 (c) 2021 IEEE. \hfill} \hspace{\columnsep}\makebox[\columnwidth]{ }}

	\maketitle
	\IEEEpubidadjcol
	\makeatletter
	\renewcommand\@afterheading{%
		\@nobreaktrue
		\everypar{%
			\if@nobreak
			\@nobreakfalse
			\clubpenalty 1
			\if@afterindent \else
			{\setbox\z@\lastbox}%
			\fi
			\else
			\clubpenalty 1
			\everypar{}%
			\fi}}
	\makeatother
	
	\begin{abstract}
		Convolutional Neural Networks (CNNs) have achieved impressive results across many super-resolution (SR) and image restoration tasks. While many such networks can upscale low-resolution (LR) images using just the raw pixel-level information, the ill-posed nature of SR can make it difficult to accurately super-resolve an image which has undergone multiple different degradations.  Additional information (metadata) describing the degradation process (such as the blur kernel applied, compression level, etc.) can guide networks to super-resolve LR images with higher fidelity to the original source.  Previous attempts at informing SR networks with degradation parameters have indeed been able to improve performance in a number of scenarios. However, due to the fully-convolutional nature of many SR networks, most of these metadata fusion methods either require a complete architectural change, or necessitate the addition of significant extra complexity.  Thus, these approaches are difficult to introduce into arbitrary SR networks without considerable design alterations.   In this paper, we introduce meta-attention, a simple mechanism which allows any SR CNN to exploit the information available in relevant degradation parameters. The mechanism functions by translating the metadata into a channel attention vector, which in turn selectively modulates the network's feature maps. Incorporating meta-attention into SR networks is straightforward, as it requires no specific type of architecture to function correctly.  Extensive testing has shown that meta-attention can consistently improve the pixel-level accuracy of state-of-the-art (SOTA) networks when provided with relevant degradation metadata.  Despite average memory/runtime overheads of less than $\approx$2.6\%/0.025 seconds for the datasets and models considered, meta-attention improves the performance for both PSNR and SSIM; for PSNR, the gain on blurred/downsampled ($\times$4) images is of 0.2969 dB (on average) and 0.3320 dB for SOTA general and face SR models, respectively.  The coding framework used for this paper is available at: \textbf{\url{https://github.com/um-dsrg/Super-Resolution-Meta-Attention-Networks}}.
	\end{abstract}
	
	\vspace{-0.15cm}
	\begin{IEEEkeywords}
		super-resolution, image restoration, convolutional neural networks, channel attention, metadata fusion.
	\end{IEEEkeywords}
	
	\IEEEpeerreviewmaketitle 
		\vspace{-0.1cm}
	\section{Introduction \& Related Work}
	\begin{figure*}[!t]
		\vspace{-0.5cm}
		\centering
		\includegraphics[width=0.8\linewidth]{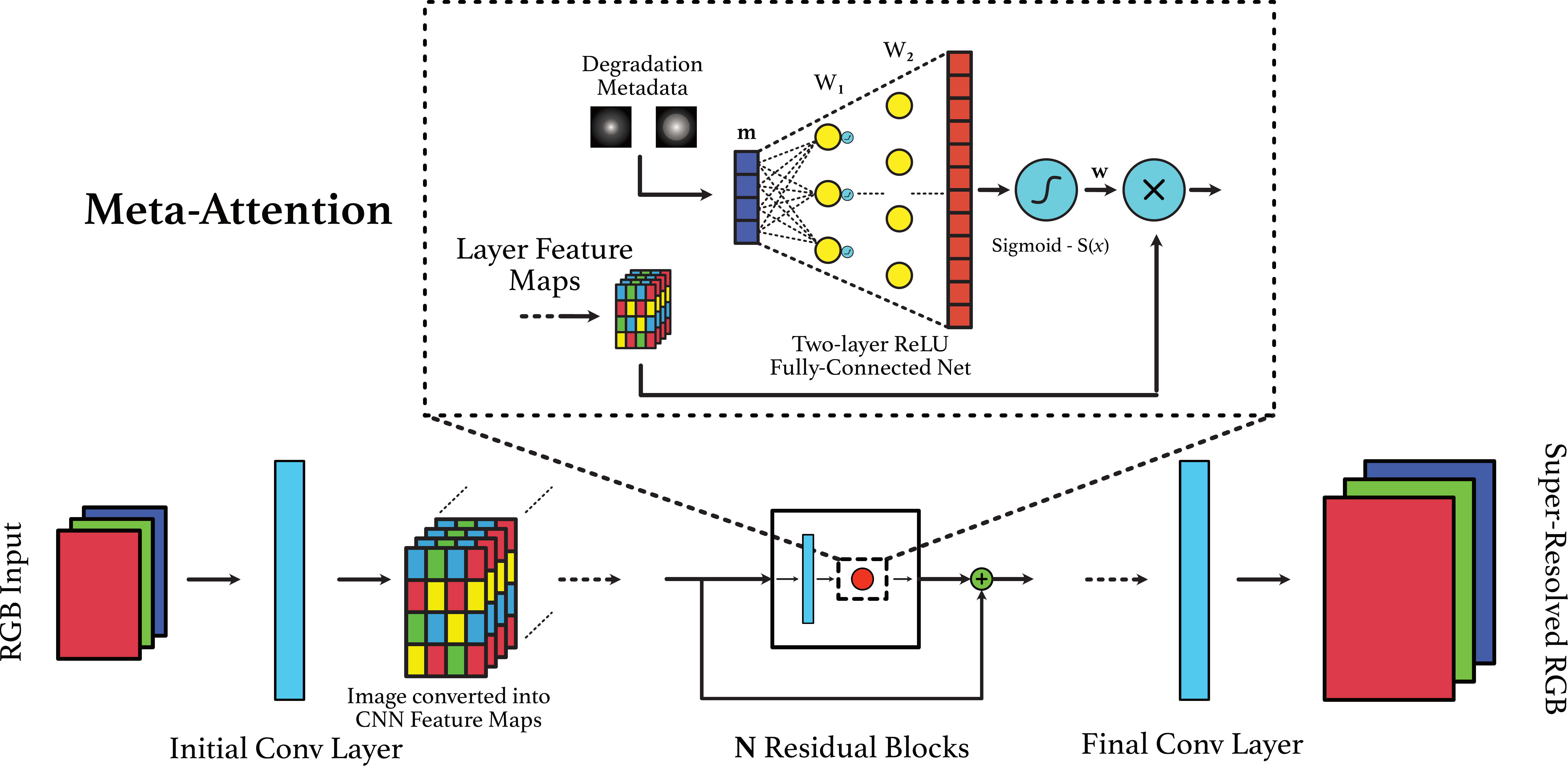}
		\vspace{-0.3cm}
		\caption{Our proposed meta-attention block is shown in the dotted box above, with the positioning in a typical residual SR network shown below.}
		\label{fig:block_diagram}
		\vspace{-0.5cm}
	\end{figure*}
	\IEEEPARstart{T}{he} task of image super-resolution (SR) is considered to be an ill-posed problem, as any given low-resolution (LR) image could be upscaled into many distinct high-resolution (HR) counterparts. Any distortions (e.g. blurring, noise injection, compression, etc.) that may affect an image, as is often the case in real-life scenarios, are typically challenging to identify correctly and thus reverse, further increasing the difficulty in yielding super-resolved images that are of satisfactory quality.  Convolutional Neural Networks (CNNs), despite being very successful at SR tasks due to their extraordinary capability for learning complex representations \cite{general_review, super_res_review}, can only extract a finite amount of information from pixel-level LR data, and thus struggle to tackle multiple degradation operations without further guidance \cite{SRMD}.
	
	However, additional information (metadata) such as image attributes or degradation parameters (e.g. the blur kernel, quantity of noise or the level of compression) could aid the SR process by narrowing down the vast HR space. Previous works have proposed CNN networks which accept blur kernels/noise levels as false image channels \cite{SRMD, SFTMD, srmd_upgrade} or as additional network inputs \cite{cornillereBlindImageSuperResolution2019, unfolding}, and thus allow their models to adapt, and respond to, the degradations used for a particular LR image. For super-resolving face images (face SR), various models have been implemented which either incorporate facial metadata (gender, age, facial features, etc.) into their networks as additional inputs \cite{supp_attributes, facial_embedding, attribute_augmented, RAAN}, or use these to construct mathematical face representations  \cite{facial_capsules}, each with reported improvements in SR quality.  Despite these successes, most attribute-fusion networks are difficult to include in generic SR CNNs. Often, attribute-fusion is proposed as either part of a stand-alone framework, or within modules that require significant design changes (e.g. placement of encoder/decoder blocks \cite{supp_attributes}) or introduce large amounts of extra complexity. Even in blind SR, where the goal is often to predict the unknown degradations used to generate an LR image, most methods do not attempt to introduce any degradation metadata into their networks \cite{ntire}, despite metadata injection having been shown to be successful in several blind SR models \cite{SFTMD,cornillereBlindImageSuperResolution2019,competing_metanetwork}. 
	
	In this paper, we propose a new mechanism for introducing metadata into SR models, hereinafter referred to as \textbf{meta-attention}.  Inspired by channel attention \cite{rcan}, meta-attention condenses image attributes into an attention vector, and uses it to specifically modulate network feature maps based on the information present in available metadata.  Lightweight fully-connected (FC) layers are used to produce the attention vector, requiring a minimal number of extra parameters to introduce meta-attention into an SR CNN.  Placement of meta-attention within a network is straightforward and can thus easily fit into any CNN architecture, without the need for extra branches or complex structural changes.  To validate this claim, we have incorporated meta-attention into a variety of state-of-the-art (SOTA) SR networks, with differing architectures.  With just $\approx$3.2\% or less extra parameters, the introduction of meta-attention has achieved an average Peak Signal-to-Noise Ratio (PSNR)/Structural Similarity Index Measure (SSIM) \cite{ssim} gain of 0.2969 dB/0.0070 over the original SR models, and a PSNR/SSIM gain of 0.3320 dB/0.0024 over a SOTA face SR model, both on blurred and downsampled ($\times$4) datasets.    
	
	The contributions of this work are thus twofold:  we present meta-attention, a technique for the introduction of metadata into any CNN network, requiring minimal changes other than the addition of meta-attention blocks within the feed-forward path of the network.  Secondly, we show that meta-attention can help improve the performance of several SOTA networks with differing architectures, across both face and general SR. 
	
	The rest of this paper is structured as follows:  Section \ref{method} discusses the implementation of the proposed meta-attention.  Section \ref{results} follows up with the technical details and results of our evaluation of various SOTA networks upgraded with meta-attention.  Finally, Section \ref{conc} presents the main conclusions drawn in this paper, and provides scope for future work.
	\vspace{-0.22cm}
	\section{Methodology} \label{method} 
	\subsection{Meta-Attention Module}\label{eqn_section}
	The proposed meta-attention mechanism is based on the concept of channel attention. Channel attention was originally conceived as a means to capture and exploit global spatial relationships in a CNN feature map, and use this to guide a network to focus on those feature maps which contain the most useful information \cite{rcan}.  The mechanism works by modulating the magnitude of each feature map based on an attention score, thus encouraging the downstream convolutional layers to focus on those feature maps having increased magnitudes.  Meta-attention attempts to achieve the same modulation effect, but instead based entirely on the provided metadata information (such as the blur kernel, noise level, and compression quantization level).  This means that the metadata is helping to steer the network towards features with increased importance for the particular degradation provided.  To achieve this effect, given metadata vector $\textbf{m}$ (details on how to obtain $\textbf{m}$ are given in Section \ref{data_setup}) for the image being super-resolved, we first pass $\textbf{m}$ through two FC layers, separated by the Rectified Linear Unit (ReLU) nonlinearity, $f(x)$.  The final layer is set up to contain as many output units as there are channels in the corresponding network.  Then, we use the sigmoid function to scale the output of each unit in the range [0, 1] to attain the attention vector $\textbf{w}$. Finally, $\textbf{w}$ is used to scale the corresponding feature maps.  The entire operation can be expressed as in Equation \ref{m-eqn}, with the corresponding block diagram provided in the dotted box of Fig. \ref{fig:block_diagram}:
	\begin{equation}\label{m-eqn}
		\textbf{w} = S(W_{2}f(W_{1}\textbf{m}))
	\end{equation}
	where $f(x)=max(0, x)$ corresponds to the ReLU function, $S(x)=1/(1+exp(-x))$ is the sigmoid function, and $W_{1}, W_{2}$ are the weights of the first and second FC layers, respectively.
	\begin{table*}[!t]
		\vspace{-0.5cm}
		\caption{SR results on blurred \& downsampled images (scale $\times$4).  Bold values refer to the best result when comparing each network to its corresponding meta-network. Runtime was averaged across all images from all test datasets.}
		\vspace{-0.7cm}
		\renewcommand{\arraystretch}{1.3}
		\begin{center}
			\begin{tabular}{ccccccccccccc}
				\hline
				\textbf{Model} & \textbf{Trainable} & \textbf{Average} & \multicolumn{2}{c}{\textbf{Set5}}  & \multicolumn{2}{c}{\textbf{Set14}}  & \multicolumn{2}{c}{\textbf{BSDS100}} & \multicolumn{2}{c}{\textbf{Manga109}} & \multicolumn{2}{c}{\textbf{Urban100}} \\
				& \textbf{Parameters} & \textbf{Runtime (s)} & PSNR & SSIM & PSNR & SSIM & PSNR & SSIM & PSNR & SSIM & PSNR & SSIM \\
				\hline
				Bicubic & - & 0.0054 &  25.1140 &                    0.7276 &                    23.3000 &                    0.6215 &                    23.6628 &                    0.5864 &                    22.0783 &                    0.7113 &                    20.7192 &                    0.5731\\
				SRMD \cite{SRMD} & 1,546,288 & 0.0289 & 29.8551 &    0.8637 &    26.3506 &     0.7425 &      25.8852 &       0.6975 &       27.9152 &        0.8731 &       23.7010 &        0.7306\\
				SFTMD \cite{SFTMD} & 4,234,691 & 0.0139 & 30.3419 &    0.8727 &    26.6556 &     0.7515 &      26.1175 &       0.7057 &       28.9294 &        0.8894 &       24.2790 &        0.7530\\
				\hdashline
				EDSR \cite{edsr} & 43,089,923 & 0.0087 &  30.2476 &    0.8741 &    26.6503 &     0.7507 &      26.1604 &       0.7059 &       28.9113 &        0.8887 &       24.3936 &        0.7564  \\
				Meta-EDSR & 44,191,747 & 0.0173 & \textbf{30.6688} &   \textbf{ 0.8776} &    \textbf{26.8903} &     \textbf{0.7579} &      \textbf{26.2681} &       \textbf{0.7116} &       \textbf{29.6419} &        \textbf{0.8986} &      \textbf{24.7200} &       \textbf{0.7695} \\
				\hdashline
				RCAN \cite{rcan} & 15,592,355 & 0.0680 &  30.3981 &    0.8744 &    26.7267 &     0.7530 &      26.1959 &       0.7072 &       29.3058 &        0.8935 &       24.6354 &        0.7646 \\
				Meta-RCAN & 16,085,155 & 0.0886 &   \textbf{30.6943} &   \textbf{0.8779} &   \textbf{26.9033} &     \textbf{0.7586} &     \textbf{26.2834} &       \textbf{0.7121} &       \textbf{29.7563} &        \textbf{0.8998} &       \textbf{24.8770} &       \textbf{0.7739}\\
				\hdashline
				HAN \cite{han} & 16,071,745 & 0.0695 &   30.5187 &    0.8762 &    26.7571 &     0.7534 &      26.1957 &       0.7068 &       29.3308 &        0.8945 &       24.6262 &        0.7641 \\
				Meta-HAN & 16,564,545 & 0.0901 & \textbf{30.7304} &    \textbf{0.8784} &    \textbf{26.9333} &    \textbf{0.7589}&      \textbf{26.2883} &      \textbf{0.7123} &       \textbf{29.7654} &        \textbf{0.9000} &      \textbf{24.8677} &        \textbf{0.7742}\\
				\hdashline
				SAN \cite{san} & 15,860,488 & 0.3853 &  30.5128 &    0.8757 &    26.6965 &     0.7515 &      26.1644 &       0.7063 &       29.1047 &        0.8900 &       24.4775 &        0.7590 \\
				Meta-SAN & 16,353,288 & 0.4615 & \textbf{30.7435} &   \textbf{0.8786} &    \textbf{26.9277} &     \textbf{0.7592} &    \textbf{26.2916}&       \textbf{0.7126} &       \textbf{29.7877} &        \textbf{0.9003}&      \textbf{24.9094} &        \textbf{0.7749}\\
				\hline
			\end{tabular}
		\end{center}
		\label{general_sr}
		\vspace{-0.7cm}
	\end{table*}
	\vspace{-0.6cm}
	\subsection{Network \& Metadata Compatibility}
	Meta-attention was designed with the goal of the mechanism being as unobtrusive and lightweight as possible, making it straightforward for inclusion within any SR CNN framework.  When adding meta-attention, module placement and quantity could be optimised for each individual network, but we opted to place a meta-attention module within all residual blocks of each network considered for the sake of consistency (as shown in Figure \ref{fig:block_diagram}).  Empirical evaluations show that this architecture is compatible with all residual networks considered, both for general and face SR (as elaborated in Section \ref{results}).
	
	Previous works have focused almost entirely on blur kernels for their meta-fusion systems.  Apart from blur kernels, we also show that our system can accept and utilise compression metadata, both alone and in tandem with blur kernel information (Section \ref{multi-meta}).  Otherwise, adding further metadata can be achieved simply by concatenating it with the 1D-vector $\textbf{m}$. 
	\vspace{-0.54cm}
	\section{Experiments \& Results}\label{results}
	\subsection{Network Training, Datasets \& Metadata Encoding}\label{data_setup}
	In order to validate our claim that meta-attention can be inserted into any SR network, and provide a tangible benefit, we inserted meta-attention into both general and face SR SOTA networks.  All models (original and modified) were trained and evaluated on the following datasets and degradations, with all results computed using an NVIDIA GeForce RTX 3090 GPU (except for bicubic interpolation, which is computed on CPU):
	
	\textbf{General SR} - DIV2K \cite{Div2k} \& Flickr2K \cite{flickr2k}: These datasets contain 800 and 2650 high-resolution training images, respectively, with varied subject matter.  We used the combined dataset (3450 images) as our training set for general SR, and applied the 100-image validation set of DIV2K for model selection.  After the best performing models were selected, test results were computed on the standard Set5 \cite{set5}, Set14 \cite{set14}, BSDS100 \cite{BSDS100}, Manga109 \cite{manga109} and Urban100 \cite{urban100} datasets. All LR images were synthesised by either of two protocols:
	\begin{itemize}[nosep,topsep=0pt, leftmargin=10pt]
		\item By first blurring and then applying bicubic downsampling with the appropriate scale factor.  Blurring was carried out using a 21$\times$21 isotropic Gaussian kernel with a width randomly selected from the range [0.2, 4] for each image, using the protocol and code in \cite{SFTMD}.  Before insertion into meta-networks, blur kernels were first downsized to a 10-dimensional vector using Principal Component Analysis (PCA) (again following the protocol in \cite{SFTMD}).  This vector becomes the input $\textbf{m}$, as described in Section \ref{eqn_section}.
		\item By blurring, downsampling (both identically as above) and then compressing the image using JM H.264 version 19.0 \cite{JM}.  Images were compressed as single-frame YUV files. For each image, a random I-slice Quantization Parameter (QPI) was selected from the range [20, 40] (based on typical values used by security cameras \cite{IPVM}).  This parameter was fed to meta-networks by first scaling it to the range [0, 1] (a QPI of 20 is scaled to 0 and a QPI of 40 is scaled to 1) and concatenating it with the blur kernel PCA vector (if this is used).  The resulting vector (containing the QPI or PCA blur kernel, or both) equates to $\textbf{m}$, as in Section \ref{eqn_section}.
	\end{itemize}
	During training, a single 64$\times$64 random patch from each LR training image was fed to an SR network in every epoch.  Random flips (vertical/horizontal) and 90$\degree$ rotations were applied for each patch.  All networks were trained from scratch using a cosine annealing scheduler \cite{cosine} and an Adam optimiser \cite{adam}. Network hyperparameters were selected according to recommendations provided by the authors. Identical hyperparameters were used for each network/meta-network pair to ensure fair comparisons.  All salient hyperparameters are available in our released code framework.  During validation/testing, entire LR images were fed to each network, unless specified otherwise. Presented test results correspond to the models with the best validation PSNR after 1300 epochs; this epoch count was selected as a compromise between performance and practicality.  
	
	We selected RCAN \cite{rcan}, EDSR \cite{edsr}, SAN \cite{san} and HAN \cite{han} as our reference SOTA networks, and constructed meta- versions of each network by inserting meta-attention blocks (ref. Section \ref{method}).  Meta-attention FC layers were initialized with the defaults for PyTorch \cite{pytorch}. 
	
	\textbf{Face SR} - CelebA-HQ \cite{celebahq}: This dataset contains 30000 1024$\times$1024 facial images, generated from the original CelebA dataset \cite{celeba}.  We extracted the images via the CelebAMask-HQ dataset \cite{CelebAMask-HQ} and used the provided CelebA train/validation/test splits ($\sim$24k images training, $\sim$3k images validation, $\sim$3k images testing) for our analysis.  HR reference images were created by first downscaling (bicubic) to 512$\times$512, to reduce GPU memory requirements during training.  The corresponding LR set (128$\times$128) was generated via blurring and downscaling of each HR image, using the same protocol as with general SR.  General SR models are capable of producing respectable results for face SR, but specialised networks which are tailored for super-resolving facial images are also available. Thus, we selected a representative general SR model, RCAN \cite{rcan}, and a SOTA face SR model, SPARNet \cite{sparnet}, to determine whether face SR can also benefit from metadata insertion.  RCAN models were trained using a cosine annealing scheduler as for general SR, while SPARNet models were trained with a fixed learning rate, as recommended in \cite{sparnet}.  Meta-RCAN has the same architecture used for general SR but is trained from scratch on CelebA-HQ images.  Meta-SPARNet was built by inserting meta-attention within each of SPARNet's residual blocks, including those within the encoder/decoder layers. Whole LR images were fed to networks during both training and testing.  Test results presented correspond to the models with the best validation PSNR after 50 epochs.
	\vspace{-0.2cm}
	\subsection{Meta-Attention in General SR}\label{results_general}
	\begin{table*}[!t]
		\vspace{-0.5cm}
		\caption{SR results on blurred, downsampled \& compressed images (scale $\times$4).  Metadata inserted is included in brackets for each model.  Bold values refer to the best performing model for each dataset/metric.}
		\renewcommand{\arraystretch}{1.3}
		\begin{center}
			\vspace{-0.2cm}
			\begin{tabular}{ccccccccccc}
				\hline
				\textbf{Model} & \multicolumn{2}{c}{\textbf{Set5}}  & \multicolumn{2}{c}{\textbf{Set14}}  & \multicolumn{2}{c}{\textbf{BSDS100}} & \multicolumn{2}{c}{\textbf{Manga109}} & \multicolumn{2}{c}{\textbf{Urban100}} \\
				& PSNR & SSIM & PSNR & SSIM & PSNR & SSIM & PSNR & SSIM & PSNR & SSIM \\
				\hline
				Bicubic &  24.2328 &    0.6741 &    22.6609 &     0.5776 &      22.7669 &       0.5356 &       21.4071 &        0.6719 &       20.1369 &        0.5299 \\
				RCAN \cite{rcan} &  26.6135 &    0.7649 &    24.4088 &     0.6437 &      23.8271 &       0.5813 &       24.5444 &        0.7778 &       21.8312 &        0.6246 \\
				Meta-RCAN (qpi) &   26.5949 &    0.7645 &    24.4318 &     0.6445 &      23.8352 &       0.5817 &       24.5945 &        0.7807 &       21.8715 &        0.6270 \\
				Meta-RCAN (blur-kernels) &  26.6018 &    0.7650 &    \textbf{24.4710} &     0.6469 &      23.8568 &       0.5837 &       24.7393 &        0.7858 &       21.9459 &        0.6320 \\
				Meta-RCAN (blur-kernels + qpi) &  \textbf{26.6553} &    \textbf{0.7664} &    24.4638 &     \textbf{0.6475} &      \textbf{23.8646} &      \textbf{ 0.5838} &       \textbf{24.7762} &        \textbf{0.7866} &       \textbf{21.9635} &        \textbf{0.6323} \\
				\hline
			\end{tabular}
		\end{center}
		\label{compressed_sr}
		\vspace{-0.5cm}
	\end{table*}
	\begin{figure*}[!t]
		\centering
		\includegraphics[width=0.88\linewidth]{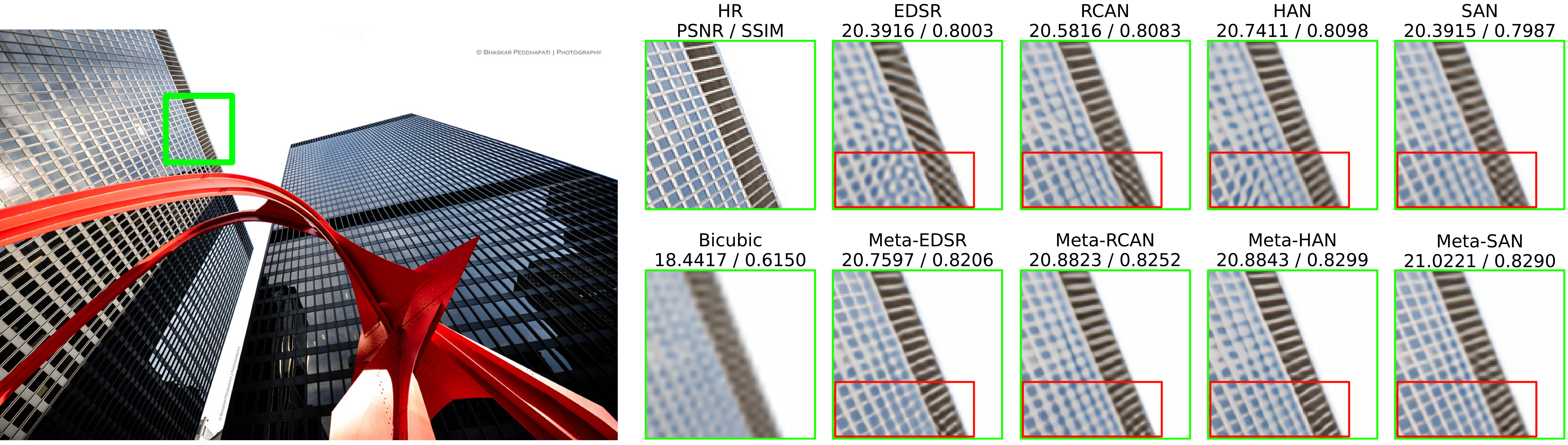}
		\vspace{-0.3cm}
		\caption{Visual comparison on img\_062 from Urban100 \cite{urban100}. LR image blurred using a 21x21 isotropic Gaussian kernel with width 0.82 and downscaled by 4.  Red boxes represent examples of areas of visual improvement achieved by introducing metadata into each model.}
		\label{fig:image_results}
		\vspace{-0.6cm}
	\end{figure*}
	Using blurred images as input and the corresponding blur kernel PCA vectors as the only source of metadata, SR test image quality was objectively measured using PSNR and SSIM \cite{ssim}, results of which are shown in Table \ref{general_sr}.  It is highly evident that the addition of blur kernel information significantly boosts the performance of each network across the board, regardless of the architecture of each baseline method.  This result is further reinforced by Figure \ref{fig:image_results}, which shows that meta-aware networks are capable of correctly super-resolving obscure patterns which are typically muddled by the original models (additional examples available in the Supplementary Material).  Furthermore, as can be observed in Table I, the overhead in terms of memory (given an average increase of approximately 2.97\%) and computational cost (with a runtime increase of approximately 0.0315 seconds per image on average) of the proposed architecture is generally marginal. It is expected that memory requirements and running times could be further reduced with optimized code.  Table \ref{general_sr} also shows how EDSR, which performs worse than SFTMD (a SOTA SR CNN specifically designed for metadata-aware SR) for several of the datasets considered, can be made to outperform SFTMD with the addition of meta-attention.
	\vspace{-0.35cm}
	\subsection{Insertion of Multiple Types of Metadata}\label{multi-meta}
	Apart from blur kernels, it can also be shown that the proposed framework is able to successfully exploit other types of metadata such as the QPI value representing the degree of image compression used.  Table \ref{compressed_sr} shows the results of adding different combinations of metadata to RCAN with meta-attention.  While adding QPI information alone to RCAN only shows weak improvements, adding both blur kernels and QPI values produces the best results, indicating that the proposed network is capable of not just handling multiple types of meta information, but also extracting and exploiting useful complementary information provided by each input.
	\vspace{-0.35cm}
	\subsection{Meta-Attention in Face SR}
	Table \ref{face_sr} shows similar trends for face SR to those observed previously, with meta-attention providing a substantial benefit to PSNR and SSIM values for both models despite an average parameter/runtime increase of just 2.16\%/0.0123 seconds.
	\begin{table}[!t]
		\vspace{-0.15cm}
		\caption{SR results on blurred \& downsampled face images (scale $\times$4).  Bold values refer to the best result when comparing a network to its corresponding meta-network.  Runtime was averaged across all images in the test dataset.}
		\vspace{-0.4cm}
		\renewcommand{\arraystretch}{1.3}
		\begin{center}
			\begin{tabular}{ccccc}
				\hline
				\textbf{Model} & \textbf{Trainable} & \textbf{Average} & \multicolumn{2}{c}{\textbf{CelebA-HQ Test Set}}    \\
				& \textbf{Parameters} & \textbf{Runtime (s)} & PSNR & SSIM \\
				\hline
				Bicubic & - & 0.0038 &  29.3012 &                    0.8115\\
				\hdashline
				RCAN \cite{rcan}  & 15,592,355 & 0.0559  & 33.1678 & 0.8815\\
				Meta-RCAN & 16,085,155 & 0.0774 &  \textbf{33.2286} &   \textbf{0.8825}\\
				\hdashline
				SPARNet \cite{sparnet} & 10,518,867 & 0.0299 &                     32.4942 &                    0.8737\\
				Meta-SPARNet & 10,641,827 &   0.0330 &                  \textbf{32.8262} &                    \textbf{0.8761}\\
				\hline
			\end{tabular}
		\end{center}
		\label{face_sr}
		\vspace{-0.7cm}
	\end{table}
	\section{Conclusions}\label{conc}
	This paper has presented meta-attention, a novel framework for introducing metadata into any SR network, regardless of architecture and application, with minimal complexity increase.  Results have shown that meta-attention provides substantial performance improvements in terms of both PSNR and SSIM on a variety of architectures, when fed with blur kernel or compression metadata (or both).  Moreover, performance gains were also observed not solely for general SR, but also for face SR which has important real-world applications such as in security and law enforcement \cite{surveillance_sr}. We envision that this framework can allow SR researchers to more easily introduce metadata into their networks, especially in applications where this could be readily available (such as extracting the QPI from video bitstreams) or in blind SR, where methods attempt to estimate the attributes of unknown degradations.  We also hypothesise that this framework could accept additional image attributes other than degradation metadata (e.g. CelebA facial features), which will be investigated as part of future work.     
	\vspace{-0.3cm}
	\section*{Acknowledgements}
	This research forms part of the Deep-FIR project, which is financed by the Malta Council for Science \& Technology (MCST), for and on behalf of the Foundation for Science \& Technology, through the FUSION: R\&I Technology Development Programme.
	\clearpage
	\bibliographystyle{IEEEtran}
	\bibliography{references}
	
\end{document}